# The Accuracy of Galileo's Observations and the Early Search for Stellar Parallax

Christopher M. Graney
Jefferson Community College
1000 Community College Drive
Louisville, Kentucky 40272
(502) 213-7292
christopher.graney@kctcs.edu
www.jefferson.kctcs.edu/faculty/graney

ABSTRACT

The question of annual stellar parallax is usually viewed as having been a "win-win situation" for seventeenth-century astronomers who subscribed to the Copernican view of universe in which the Earth orbits the Sun and the Sun is one of many suns (the fixed stars) scattered throughout space. Detecting parallax would be solid evidence for the Earth's motion, but failure to detect parallax could be explained by the stars lying at great distances. Recent work pertaining to Galileo's observations of double stars illustrates Galileo's skill as an observer. It also indicates that, given the knowledge of optics of the time, Galileo could expect his measurements to be accurate enough that they would have revealed stellar parallax had it existed. Thus parallax was not a "win-win situation" after all. It could be solid evidence against the Earth's motion, evidence which fortunately did not dissuade Galileo from the Copernican view.

# I. INTRODUCTION

For the greater part of the five centuries that have elapsed since Copernicus proposed a heliocentric model of the universe in his *De Revolutionibus Orbium Coellestium*, astronomers who subscribed to Copernicus's view had to deal with the issue of annual stellar parallax.  Until stellar parallax was actually observed roughly two centuries ago, heliocentrists had to argue that, while they believed the universe was populated with stars whose distances from Earth's Sun varied and therefore stellar parallax should exist, lack of observable parallax did not disprove the heliocentric model.  After all, if the stars are sufficiently distant the stellar parallax would exist but simply be too small to measure.  Copernicus himself stated that the lack of any parallax meant that the stars lay at vast distances beyond the planets.[1]  Stellar parallax was a "win-win situation" for heliocentrists – failure to detect it proved nothing, while a successful detection would be powerful evidence for heliocentrism.[2]

---

[1] Copernicus, *On the Revolutions of Heavenly Spheres* [*De rev.*], translated by Charles Glen Wallis (Prometheus Books, Amherst, New York, 1995).  After describing his heliocentric model, Copernicus states "…now the careful observer can note why progression and retrogradation appear greater in Jupiter than in Saturn and smaller than in Mars….  And why these reciprocal events appear more often in Saturn than in Jupiter, and even less often than in Mars….  In addition, why when Saturn, Jupiter, and Mars are in opposition they are nearer to the Earth than at the time of their occultation and their reappearance….  All these things proceed from the same cause, which resides in the movement of the Earth."  In other words, retrograde motion and varying brightness in the planets is essentially a manifestation of Earth's motion – it is a planetary parallax akin to annual stellar parallax.  Copernicus then continues "But that there are no such appearances among the fixed stars argues that they are at an immense height away, which makes the circle of annual movement or its image disappear from before our eyes since every visible thing has a certain distance beyond which it is no longer seen, as in optics.  For the brilliance of their lights shows that there is a very great distance between Saturn the highest of the planets and the sphere of the fixed stars.  By this mark in particular they are distinguished from the planets, as it is proper to have the greatest difference between the moved and the unmoved.  How exceedingly fine is the godlike work of the Best and Greatest Artist! [pp. 26-27]"

[2] Harald Siebert, "The Early Search for Stellar Parallax:  Galileo, Castelli, and Ramponi", *Journal for the History of Astronomy*, Vol. 36 (2005), 251-271.  Siebert writes "Thus Copernicans were actually placed in a win-win situation to resolve the dispute over the true world system.  They had in hand the 'experimentum crucis' of the debate, and Galileo himself, as he tells us in his *Dialogo*, was fully aware of this unequal balance…. [pp. 251-2] "

Recent work has brought to light evidence that as early as the 1610's Galileo Galilei and Benedetto Castelli were actively searching for stellar parallax using the newly developed telescope.[3,4]  However, the assumption in this recent work remains that parallax could not disprove heliocentrism – only prove it.  This assumption may not be correct.  In this paper I will argue the following:

- Galileo's skill as instrument-builder and observer was such that Galileo recorded observations with arc-second accuracy.

- Such accuracy meant that Galileo could obtain reproducible measurements of stellar sizes and distances that, while today we know them to be erroneous, would be viewed as reliable given the knowledge of his time.

- Those measurements would mean that stellar parallax could and, given the knowledge of his time, did "disprove" heliocentrism.

At the end of the paper I will discuss questions for future study relating to Galileo's fortuitous decision to support heliocentrism despite what his observations told him.

## II.  GALILEO'S SKILL AND ACCURACY

There are many reasons for believing that Galileo could make and record observations with arc-second accuracy.  For example, Galileo's observations of Jupiter have been shown to be of high quality – he recorded the separations between Jupiter and its moons to within 0.1 Jovian radii; in his sketches he places Jupiter's moons to an accuracy of less than the width of the dots he uses to mark the moons' positions; he recorded positions of

---

[3] Leos Ondra, "A New View of Mizar", *Sky and Telescope*, July 2004, 72-75. p. 73.
[4] Siebert, *JHA* (ref. 2), p. 257.

objects as faint as Neptune.[5]  An interesting illustration of the accuracy of Galileo's work

can be seen by comparing some of his sketches of the Jovian system to a simulated

telescopic view generated by planetarium software [Figure 1].  This accuracy is also on

display in a sketch Galileo made on 4 February 1617 of a grouping of stars in Orion

(including three stars in the Trapezium).[6]  A comparison of that sketch to modern data on those stars again reveals Galileo's skill and the quality of his instruments – Galileo was able to produce an accurate sketch of stars that were separated by less than 15 arc-seconds [Figure 2].

By 1612 Galileo had stated that he was able to make measurements accurate to

within arc-seconds,[7] and his statements about and measurements of stars are consistent

with that claim.  In his *Dialogue Concerning the Two Chief World Systems* (1632)

Galileo states that a first-magnitude star has a diameter of 5 arc-seconds while a sixth-magnitude star has a diameter of one-sixth that size (50/60 arc-second or 0.83 arc-second).[8]  He goes on to calculate the distance to a sixth-magnitude star (2160 times the distance to the Sun, or 2160 A.U.) based on the assumption that a star is equal to the Sun in actual size and that its apparent size is determined by distance and geometry.[9] This implies that stellar magnitudes and sizes are linearly related:

| Magnitude | Stellar Diameter (arc-seconds) |
|-----------|-------------------------------|
| first | 300/60 or 5.00 |
| second | 250/60 or 4.17 |
| third | 200/60 or 3.33 |
| fourth | 150/60 or 2.50 |
| fifth | 100/60 or 1.67 |
| sixth | 50/60 or 0.83 |

Today we know that the apparent size of a star as viewed through a small but optically fine telescope[10] is largely a function of wave optics and does not reflect the physical size of the star. However, a linear relationship between magnitude, size, and distance would have been a defensible proposition for an observer such as Galileo who knew nothing of wave optics [Figure 3]. Galileo believed that a good telescope did not produce illusions – that it accurately showed the true sizes of stars.[11]

Galileo's recorded observations of stellar sizes agree with the values given above, although not perfectly. For example, undated notes of Galileo's show that he observed Sirius and measured its diameter to be 5 and 18/60 arc-seconds.[12] This is consistent with the sizes implied in the *Dialogue*. Galileo also observed the star Mizar and found it to consist of two component stars, the brighter of which (A) he measured to have a radius of 3 arc-seconds, and the fainter of which (B) he measured to have a radius of 2 arc-

---

seconds.[13,14]  This undated observation was probably made on 15 January 1617.[15]  Mizar A is about magnitude 2 and B is about magnitude 4, so the diameters of 6 and 4 arc-seconds that Galileo determined in 1617 differ from the sizes implied in the *Dialogue* fifteen years later, but by less than two arc-seconds.  Galileo measured the center-to-center separation between the two stars to be 15 arc-seconds.  Galileo's separation measurement is within half an arc-second of modern measurements,[16] and his radius measurements agree closely in proportions to what would be expected for optically good telescopes of the sizes he used [Figure 4].

In short, Galileo's notes and writings indicate that he was able to make and record observations to a high level of accuracy.  That Galileo achieved such results with some of the first telescopes known to be used to study the heavens, especially when some of his contemporaries were unable to see even relatively easy celestial sights with their telescopes[17], is a testament to his talent and work ethic.[18]

## III. STELLAR PARALLAX – NOT A "WIN-WIN" SITUATION

Unless the previously mentioned results were all flukes, and his claim of arc-seconds accuracy an exaggeration, Galileo must have been able to produce similar-quality observations at other times.  The quality of Galileo's observations now comes to bear on the question of parallax and whether it was a "win-win situation" for heliocentrists.

---

[13] Ondra, *S&T* (ref. 3), p. 74. Ondra's article includes a copy of Galileo's original notes on Mizar.

[14] Galileo, *Opere*, Vol. 3 part 2, p. 877.  Galileo's Mizar notes in the *Opere*.

[15] Siebert, *JHA* (ref. 2), p. 259.

[16] Siebert assumes that the closeness of Galileo's value to the modern value is an accident (Siebert, *JHA* (ref. 2), p. 259).

[17] Albert Van Helden, "The Telescope in the Seventeenth Century", *Isis*, Vol. 65 (1974) 38-58.  pp. 43-44, 52.  Van Helden notes one observer in 1612, Jacob Christmann, describing Jupiter as three or four fiery balls with hairs coming out like a comet. This was at the same time that Galileo was noting that he had improved his observations to an accuracy of a few arc-seconds.

[18] For another testament to Galileo's talent, and a further demonstration of his accuracy and of what he was able to achieve with his telescopes, the reader is advised to study Tom Pope and Jim Mosher's web site, "CCD Images from a Galilean Telescope", (www.pacifier.com/~tpope).  Pope and Mosher constructed a Galilean telescope and obtained afocal CCD images through it.  By comparing their CCD images with Galileo's notes and sketches, they too find Galileo to be remarkably accurate in his observations.  Pope and Mosher's work served as the inspiration for a large part of this paper.

Galileo used his observation of Mizar to calculate that Mizar A, being 1/300 the apparent radius of the Sun, must be 300 times more distant than the Sun (300 A.U.).[19]  In doing this Galileo assumed that stars are suns.  By the same logic, Mizar B would be 450 A.U. distant.

Today we know these distances to be inaccurate in the extreme.  We know that the differing stellar radii Galileo was measuring represented nothing more than a combination of a wave optics diffraction pattern/Airy Disk and the limits of the human eye [Figures 3 and 4].  However, as mentioned earlier in this paper, he could not know this.  He thought he was seeing the physical bodies of stars.  For all other objects Galileo observed, size was inversely proportional to distance.  He used size arguments to argue, for example, that Mars and Venus orbited the Sun.[20]  An understanding of wave optics and Airy Disks was far in the future when Galileo was using his telescope.  There was no reason for him to think that the size of a star was any less related to geometry and distance than was the size of Mars or Venus.

---

[19] Galileo, *Opere*, Vol. 3 part 2, p. 877.

[20] Galileo, *Dialogue* (ref. 8), "SIMP. How do you deduce that it is not the earth, but the sun, which is the center of the revolutions of the planets?  SALV. This is deduced from the most obvious and therefore most powerfully convincing observations.  The most palpable of these, which excludes the earth from the center and places the sun there, is that we find all the planets closer to the earth at one time and further from it at another.  The differences are so great that Venus, for example, is six times as distant from us at its farthest as at its closest, and Mars soars nearly eight times as high in the one state as in the other….  SIMP. But what are the signs that they move around the sun?  SALV. This is reasoned out from finding the three outer planets – Mars, Jupiter, and Saturn – always quite close to the earth when they are in opposition to the sun, and very distant when they are in conjunction with it.  This approach and recession is of such moment that Mars when close looks sixty times as large as when it is most distant. [pp. 321-322]"  When Galileo says sixty times he is referring to size in terms of area, which corresponds to a variation of just under eight times in terms of diameter or radius.  Galileo restates all this later: "SALV. …if it were true that the distances of Mars from the earth varied as much from minimum to maximum as twice the distance from the earth to the sun, then when it is closest to us its disk would have to look sixty times as large as when it is most distant….[p. 334]"  Galileo goes on to say that "…both Mars and Venus do show themselves variable in the assigned proportions…." but that these variations are only visible by means of the telescope (p. 335).

At distances of 450 A.U. and 300 A.U., Mizar's two components should have had parallax angles of 7.6 arc-minutes and 11.5 arc-minutes respectively.[21]  Over the course of a year, this would mean that the separation of the two would vary by several arc-minutes.  Since they were separated by only 15 arc-seconds and since Galileo could observe with arc-second accuracy, the Earth's motion should have revealed itself easily after a short time.

A similar situation existed in the Trapezium.  Here the stars again have separations of almost 15 arc-seconds, but in this case their sizes as recorded by Galileo varied by a factor of 4 or 5, rather than by a factor of 1.5 as in the case of Mizar.[22]  This of course would translate into their distances varying by a factor of 4 or 5 under the assumptions that stars were suns.  Even if we assume that the brightest of the Trapezium stars (Galileo's star "g" in Figure 2 – a fifth magnitude star) was an order of magnitude more distant than Mizar A (and thus substantially more distant than even the distance of sixth magnitude stars that Galileo would later discuss in his *Dialogue*), the difference in their parallax angles is such that their separation should show variations over the course of a year on the order of an arc-minute.[23]

Of course such parallax-produced swings in the separations of Mizar and the Trapezium do not occur.[24]  Given that Galileo's measurements and calculations were perfectly logical within the knowledge of his time, the lack of observable parallax would indicate that one of the assumptions involved in the observations and calculations was wrong. Either the Earth was not moving, or the stars were not suns at differing distances from Earth.[25]

---

[21] $\tan^{-1}(1/300) = 0.191^o = 11.5' = 688''$.  $\tan^{-1}(1/450) = 0.127^o = 7.6' = 458''$.

[22] Galileo, *Opere*, Vol. 3 part 2, p. 880.

[23] If star g [Figure 2] lies at 3000 A.U. and its fainter companion is only four times as distant (12,000 A.U.): $\tan^{-1}(1/3000) = 0.0191^o = 1.15' = 69''$.  $\tan^{-1}(1/12,000) = 0.00478^o = 0.29' = 17''$.

[24] Siebert argues that Galileo was aware of other close star groupings besides Mizar and the Trapezium. Parallax was not seen in them, either (Siebert, *JHA* (ref. 2), pp. 258-262).

If the Earth is not moving, then heliocentrism is invalid.

If the stars are not identical to the Sun, but rather vary in size so that, for example, Mizar A is Sun-like but Mizar B is smaller than the Sun, that could explain the lack of observable parallax in the two stars. But this would require the stars to essentially lie on a sphere that is centered on the Sun. If B was as much as two percent more distant than A Galileo could have detected parallax-induced variations in the A-B separation.[26] If we operate on the assumption that the stars are much further away, as we did with the Trapezium, we find Galileo could have detected variations if stars differed by thirty percent in distance.[27]  The result in either case is that the heavens would be a great shell of stars centered on the Sun.[28]

---

[25] We could argue that it never occurred to Galileo to look for stellar parallax in close groupings of stars, as I know of no Galileo notes explicitly stating "I have been observing Mizar and the Trapezium for months now and there is no sign of parallax." However, Siebert presents a convincing case that Galileo had to be aware that close star groupings could reveal parallax, citing a letter from one Lodovico Ramponi to Galileo outlining the idea in 1611 (Siebert, *JHA* (ref. 2), p. 254). Ramponi's letter can be found in the *Opere* (Vol. 11, pp. 159-162). If we argue that Galileo never once looked for parallax in the close star groupings he found, then we must explain why he would have declined to do so when his measurements and calculations would have lead him to believe the effect would be easy to detect.

[26] Following the angle calculations used in previous notes: If Mizar A lies at distance $s$ from the Sun (in A.U.), if B lies $k$ times further than A, and if Galileo can detect variations in the stars' separations of amount $\Delta$, then to a reasonable approximation $\Delta = tan^{-1}(1/s) - tan^{-1}(1/(ks))$, and $k = \{s[tan(tan^{-1}(1/s)-\Delta]\}^{-1}$. For Mizar A, $s = 300$ A.U. Galileo measured Mizar's components to be separated by 15 arc-seconds. Let us very conservatively assume that Galileo can detect variations in separations on the order of this value so $\Delta = 15$ arc-sec. These values yield $k = 1.022$; Galileo could detect relative changes in the separation of Mizar A and B due to parallax if B was 2.2% more distant than A. If we are less conservative and say that Galileo could detect changes comparable to the radii of the stars that he measured, so that $\Delta = 3$ arc-sec, then $k = 1.004$.

      Another possibility was that the stars were not like the Sun at all.  If they were far vaster than the Sun they could lie at even vaster distances than we've calculated so far and the question of stellar parallax would again be a "win-win situation".  If Mizar A was more than an order of magnitude larger than the Sun, and thus at a distance of 6,000 A.U. instead of 300 A.U., then the parallax might be difficult to detect.[29]  The heavens in this case would indeed be huge, but the heavens would be populated by stars that were twenty times the size of the Sun.  Instead of being similar to the Sun, the stars would be an unknown type of object.[30]

IV. QUESTIONS

Despite all this, Galileo apparently chose to stand by the view that heliocentrism was correct, with the stars in fact suns at varying distances from Earth.[31]  He even rather

---

[27] Using the 3000 A.U. value for star "g" [Figure 2] and $\Delta$ = *15 arc-sec*, *k = 1.279*.  Using $\Delta$ = *3 arc-sec*, *k = 1.046*.

[28] Another possibility would be that stars could be of differing sizes and grouped, so that Mizar A and star "g" [Figure 2] lay at differing distances, but had smaller companions with them, so that close star groupings were not just chance line-of-sight alignments.  This idea lay well in the future.

[29] If Mizar A lay at 6000 A.U., B would lie at 9000 A.U. $\tan^{-1}(1/6000) = 0.0095^\circ = 0.57' = 34"$.  $\tan^{-1}(1/9000) = 0.0064^\circ = .38' = 23"$.  This would be hard to detect under the $\Delta$ = *15 arc-sec* criterion stated in notes above.  If the $\Delta$ = *3 arc-sec* criterion is used then the stars would be all that much further and larger.

[30] One possibility that has been raised in discussions of this topic with colleagues has been that perhaps the lack of parallax could be explained if stars were suns but simply did gave off differing amounts of light.  This possibility is not valid, however.  Galileo, not being aware of Airy Disks, etc. would have assumed he was seeing the full disk of a star, much as he might see the full disk of Mars or Jupiter.  As an example, suppose two stars of the Sun's size were at the same distance but gave off differing amounts of light.  These two stars would appear equal in apparent radius but of differing surface brightness.

[31] Galileo, *Dialogue* (ref. 8).  Salviati states that the stars are suns: "See, then, how neatly the precipitous motion of each twenty-four hours is taken away from the universe, and how the fixed stars (which are so many suns) agree with our sun in enjoying perpetual rest [p. 327]."  Salviati uses the assumption that stars are suns in calculating the distance to a sixth magnitude star (p. 359).  This calculation is consistent with his Mizar calculations.  He later states "…I do not believe that the stars are spread over a spherical

surprisingly proposed in the *Dialogue* that a close pairing of stars of different brightness could yield proof of Earth's motion – as though such observations had never been tried.[32]

Since Galileo was right in his views – the Earth does orbit the Sun and the stars are suns at varying distance from Earth – this apparently is a very fortuitous case of someone standing by a theory against contrary observational evidence.  We can imagine the history of science had Galileo been swayed by his data into questioning his views and concluding that either the Earth was not moving and perhaps Tycho Brahe was right, or that the Sun was at the center of a universe bounded by a shell stars, or that the Sun was a unique body amid a vast universe of vast and alien stars.  Had Galileo put his forceful style of argument behind any of these propositions, it seems likely that a correct understanding of the nature of the universe would have been delayed by a significant amount of time.

We can also imagine the history of science had Galileo not been particularly swayed by his lack of success in detecting stellar parallax, but still decided to publish his results anyway as interesting data regarding an important scientific question.  Galileo may not have been swayed from his views by his data, but it seems safe to say that others might have been swayed had they been given the chance to see the data.  No doubt the geocentrists of Galileo's time would have realized that stellar parallax was not a "win-win situation" for heliocentrists after all.  They would have claimed that the data undermined Copernicus and probably would have hunted for more close star pairings as further support for a geocentric view.  Again, it seems likely that a correct understanding of the nature of the universe would have been delayed by a significant amount of time.

Lastly, it is not too speculative to suppose had the geocentrists known when the *Dialogue* appeared that Galileo's double-star data had existed for over a decade, unpublished, while Galileo advocated in the *Dialogue* views that the data argued against, they would not have approved.  Whatever insights motivated Galileo to personally remain convinced of heliocentrism despite the data would probably not have impressed his opponents as much as the data.  They would have felt Galileo was withholding information that undermined his view of the universe.

V.  CONCLUSION

The quality of Galileo's instruments and observations meant that stellar parallax was not a "win-win situation" for heliocentrists after all.  Careful observations combined with the limited understanding of optics of the time would yield results that challenged the view that the Earth circled the Sun, which in turn was one of many suns at scattered throughout the universe.  Fortunately for the progress of science, Galileo did not share all his observations with the world at large.  Scholars who have an interest in this topic can find much to explore regarding questions about the early search for stellar parallax, how that search supported or undermined the Copernican heliocentric model, why Galileo made the decisions he made about supporting heliocentrism and about which results to share with the world at large, and how Galileo's parallax work would have been received by his opponents.

**Figure 1**

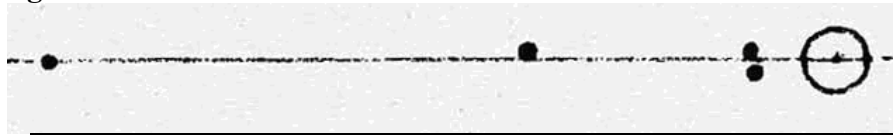

Galileo
25 March 1613 H 0.5
(*Opere* Vol. 5 p. 241)

*Stellarium*
25 March 1613
12:56:00 EST
FOV 0.367º

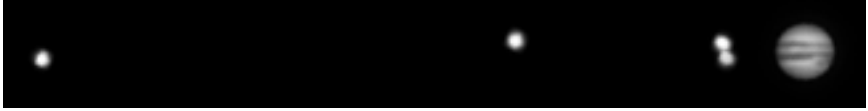

Galileo
12 March 1613 H 5
(*Opere* Vol. 5 p. 241)

*Stellarium*
12 March 1613
4:52:00 PM EST
FOV 0.366º

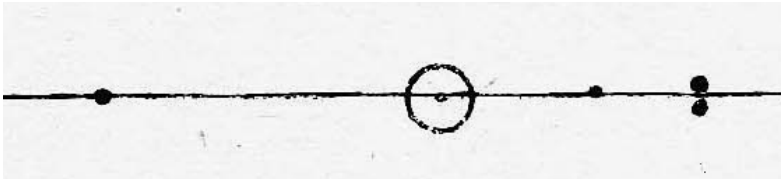

Galileo
29 March 1613
H 0.0.30
(*Opere* Vol. 5 p. 243)

*Stellarium*
29 March 1617
12:52:00 PM EST
FOV 0.368º

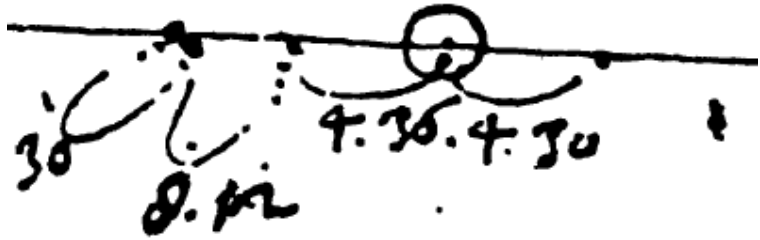

Galileo
6 January 1613
(Standish & Nobili)

*Stellarium*
6 January 1613
12:02:58 AM EST
FOV 0.368º

**Figure 1 caption:**
Galileo's drawings compared to output from the *Stellarium* open-source planetarium
software (www.stellarium.org). 6 January 1613 shows Neptune in the lower-right corner.
*Stellarium* shows Neptune as being magnitude 7.9 at the time. The *Stellarium User Guide*
states that positions of Galilean satellites are valid for 500 A.D. - 3500 A.D. but does not
state a level of precision. It states that the positions of Jupiter and Neptune are accurate to
1 arc-second. Differences exist between sketches and Stellarium due to Galileo's errors,

***Stellarium's*** **errors, and the author's limits in estimating the moment in time a given sketch represents, but the comparison is remarkable nonetheless.**

**Figure 2**

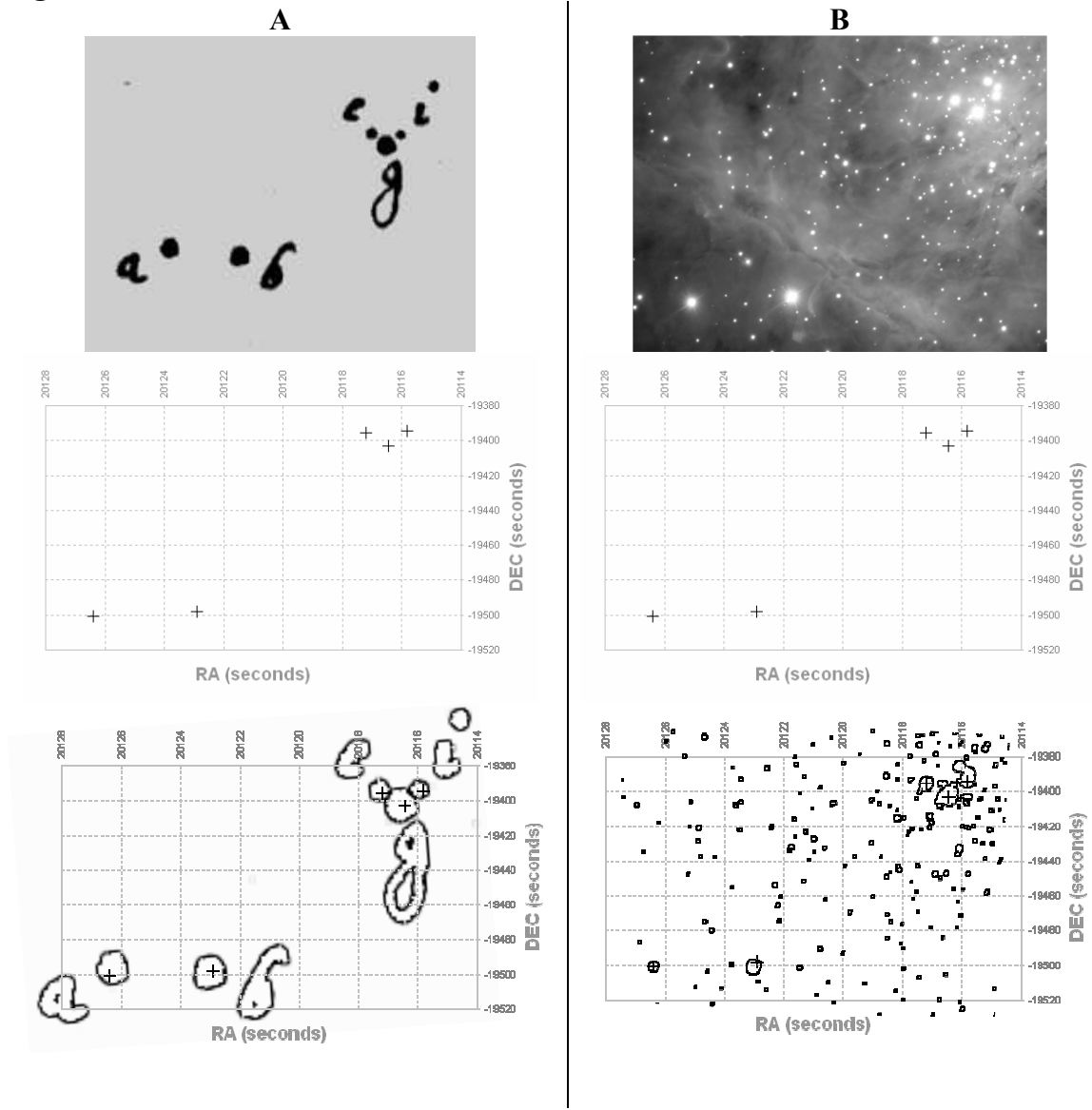

Figure 2 caption:
**Column A (top row) is Galileo's sketch of five stars in Orion which Galileo labeled a, b, c, g, i. The second row is a plot of the locations of stars HD 37042, HD 37041, HD 37023, HD 37022, and HD 37020 plotted according to their ICRS 2000.0 coordinates as given by the SIMBAD astronomical database (http://simbad.harvard.edu/Simbad). According to the SIMBAD database, none of the stars in question have large enough proper motions to produce substantial changes between 1617 and 2006. The third row superimposes these two, with Galileo's sketch processed to show his markings as white areas circled by a black border, and rotated and enlarged to match the plot. Column B features the same method applied to a January 2001 image of the Trapezium from the European Southern Observatory (http://www.eso.org/outreach/gallery/vlt/images/Top20/Top20/press-rel/phot-03a-01-normal.jpg) for the purpose of comparison, as well as of illustrating limits in the author's software and skills in the area of image processing and alignment.**

**Figure 3**

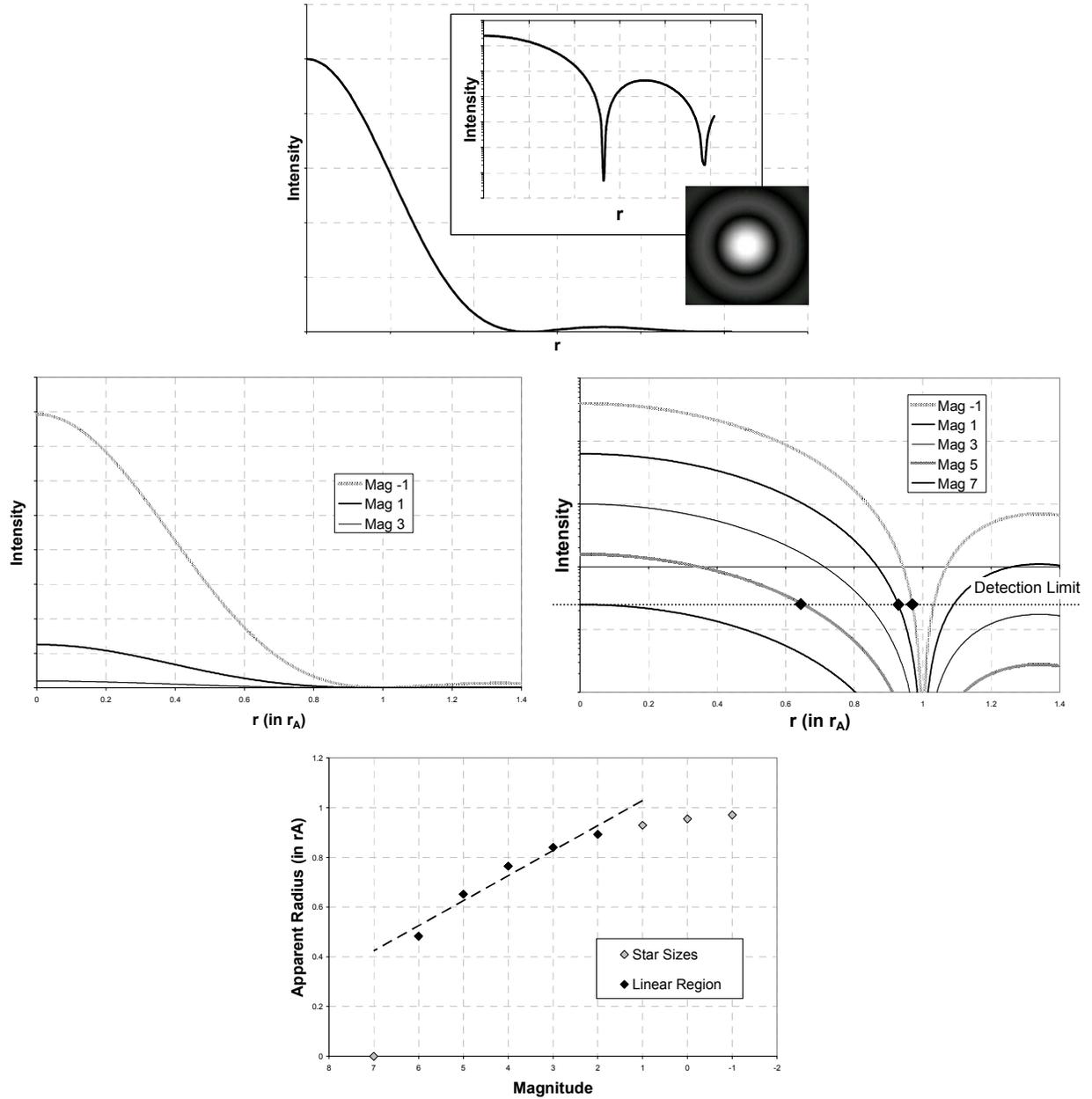

Figure 3 caption:

**The image of a star formed by a telescope is a diffraction pattern consisting of a central maximum (Airy Disk) whose angular radius is given by $r_A = 1.22\lambda/D$. Here $\lambda$ is the wavelength of light (taken in this paper as 550 nm, the center of the visible spectrum) and D is the telescope's aperture. The intensity in the pattern as a function of radius is given by $I=I_0[J_1(r)/r]^2$ where $J_1(r)$ is a Bessel function of the first kind. In the top row we see plots of I vs. r along with an illustration of the diffraction pattern. The plots of I vs. r include both a linear axis plot with which most readers will be most familiar, and a semi-log plot (insert) which will be more useful for the purposes of this paper. While the images of all stars have the same Airy Disk radius (middle row, left), they do not all have the same intensities.**

Consider a telescope system (consisting of telescope, eye, and sky conditions) which can detect stars of magnitude 6 but barely lacks the sensitivity to detect stars of magnitude 7. Thus there is an intensity limit below which the eye detects nothing, and above which the eye detects starlight. This limit is shown as a horizontal line on the semi-log plot in the middle row (right), just above the intensity curve for a seventh magnitude star. The result of this limit is that the stars will have differing sized apparent radii, shown by the diamond-shaped markers on the plot where the stars' intensities drop below the limit. Thus a fifth magnitude star will have an apparent radius of about 0.65 $r_A$, while a first magnitude star will have an apparent radius of 0.93 $r_A$ and a negative-first magnitude star will have an apparent radius nearly equal to the Airy Disk radius. A plot of apparent radius vs. magnitude is shown in the bottom row. Note that for middling magnitude stars, the relationship would appear essentially linear to observers. This translates into a linear relationship between magnitude and distance if we assume that all stars are roughly equal in actual size and that their apparent size is a function of distance and geometry as is true for every other object. This is especially true considering that truly bright stars that break from the line on the graph are comparably few in number and that faint stars that break from the line are a challenge to observe and measure. For a larger telescope on a clear night both the limiting magnitude and $r_A$ would be smaller and the "brighter is bigger" linear relationship would be less obvious. For a smaller telescope the relationship would be more pronounced. But for a sufficiently skilled observer using a single small telescope or similar small telescopes, the idea that a sixth magnitude star is six times as distant as a first magnitude star was not just a convenient assumption but a defensible deduction.

**Figure 4**

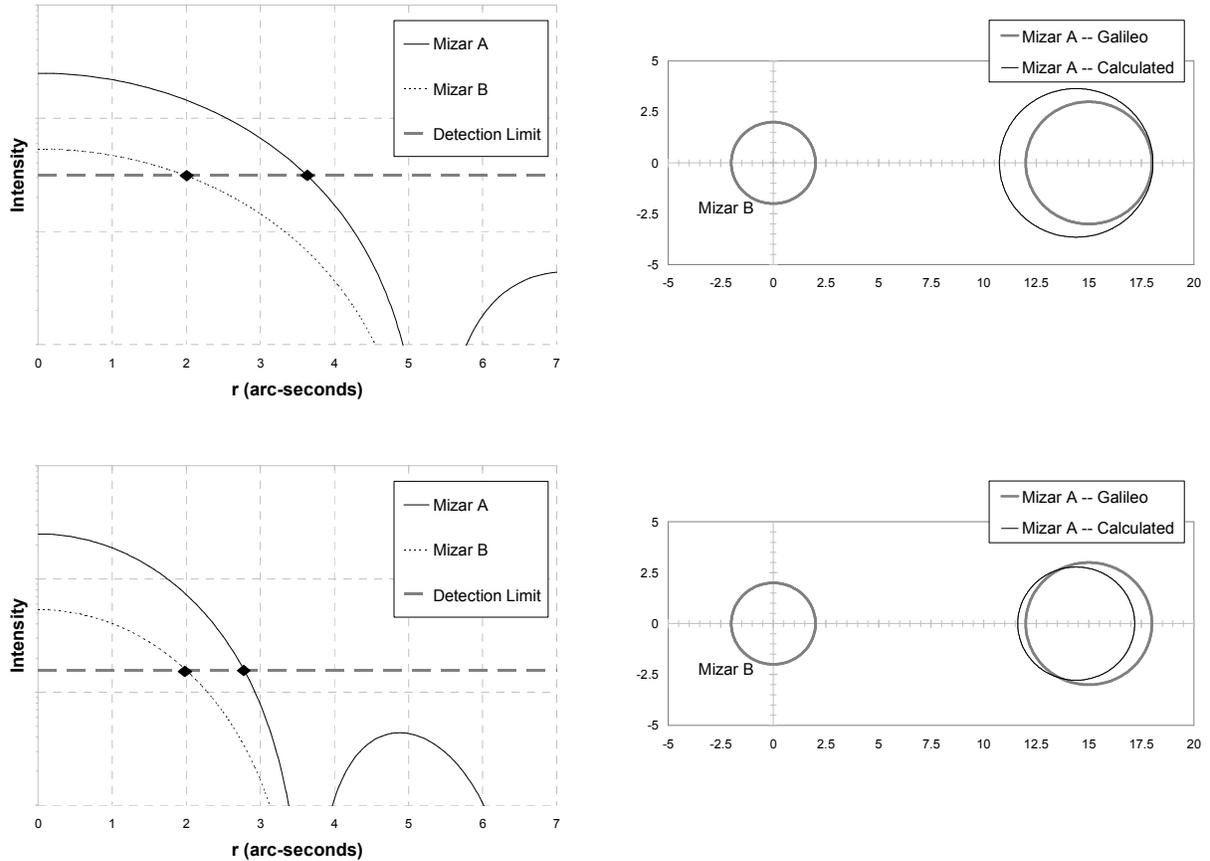

Figure 4 caption:

**The magnitudes of Mizar A and B (HD 116656 and HD 116657) according to the SIMBAD database are 2.27 and 3.95 respectively. Their relative motions are not significant enough to greatly alter their separation between 1617 and 2006. Their separation according to Hipparcos data from the Millennium star atlas (http://www.rssd.esa.int/Hipparcos/msa-tab7.html) is 14.4 arc-seconds. Galileo observed A to have a radius of 3 arc-seconds and B to have a radius of 2 arc-seconds, with a center-to-center separation of 15 arc-seconds. Plotting the intensity curves for these two stars based on a 26 mm telescope (top left) and setting a detection level such that B will have a 2 arc-second radius yields an expected radius for A of 3.65 arc-seconds (shown by the diamond-shaped markers). At top right we see a diagram of Mizar as Galileo measured it and as we would expect it to appear today via calculations for a 26 mm telescope system whose limit of detection gives B a 2 arc-second radius. In the bottom row we see the same calculations done for a 38 mm telescope. Regardless of the telescope size used, the agreement between what Galileo observed and the results of the calculations is remarkable. The 26 mm and 38 mm apertures correspond to telescopes that are attributed to Galileo.**

19